\documentstyle[twoside,11pt]{article}
\begin{document}

\title{Spherically symmetric collapse of a perfect fluid in f(R) gravity}

\author{Soumya Chakrabarti\footnote{email: adhpagla@iiserkol.ac.in}~~~and~~Narayan Banerjee\footnote{email: narayan@iiserkol.ac.in}
 \\
Department of Physical Sciences, \\Indian Institute of Science Education and Research, Kolkata\\ Mohanpur Campus, Nadia, West Bengal 741246, India.
\date{}
}

\maketitle
\vspace{0.5cm}
{\em PACS Nos. 04.50.Kd; 04.70.Bw
\par Keywords : gravitational collapse, f(R) gravity, singularity, horizon}
\vspace{0.5cm}

\pagestyle{myheadings}
\newcommand{\be}{\begin{equation}}
\newcommand{\ee}{\end{equation}}
\newcommand{\bea}{\begin{eqnarray}}
\newcommand{\eea}{\end{eqnarray}}

\begin{abstract}
The present work investigates the gravitational collapse of a perfect fluid in $f(R)$ gravity models. For a general $f(R)$ theory, it is shown analytically that a collapse is quite possible. The singularity formed as a result of the collapse is found to be a curvature singularity of shell focusing type. The possibility of the formation of an apparent horizon hiding the central singularity depends on the initial conditions.
\end{abstract}

\section{Introduction}
Although general relativity appears to be the best theory of gravity without much of a contest, various modifications have been attempted through the years for various purpose. Brans-Dicke theory\cite{brans} was proposed in order to accommodate Mach's principle in a relativistic theory of gravity. Replacing the Ricci curvature $R$ by any analytic function of $R$ in the Einstein-Hilbert action is generically called an $f(R)$ theory of gravity. Actually every different function $f=f(R)$ leads to a different theory. Besides the explanation ``why not?'', the primary motivation was to check whether such an $f(R)$ theory, particularly for $f(R) = R^{2}$, can give rise to an inflationary regime in the early universe\cite{staro, kerner}. The implications of an $f(R)$ theory in the context of cosmology was investigated by Barrow and Ottewill\cite{barrow1}. In the context of the discovery that the universe is undergoing an accelerated expansion at the present epoch, $f(R)$ theories fin
 d a rejuvenated interest so as to provide a possibility of a curvature driven acceleration where no exotic matter component has to be put in by hand. After the intial work by Capozziello {\it et al}\cite{capo} and Caroll {\it et al}\cite{caroll}, a lot of work has been done where a late time acceleration of the universe has been sought out of inverse powers of $R$ in the Einstein-Hilbert action. Das, Banerjee and Dadhich\cite{sudipta} showed that it is quite possible for an $f(R)$ gravity to drive a smooth transition from a decelerated to an accelerated phase of expansion at a recent past. The various $f(R)$ theories and their suitability in connection with various observations have been dealt with in detail by Amendola {\it et al}\cite{amendola1, amendola2}, Felice and Tsujikawa\cite{felice}, Nojiri and Odintsov\cite{nojiri}. \\

Most of the $f(R)$ theories suffer from one drawback, they fail  to match General Relativity so far as the explanation of local astronomical tests are concerned. For instance, it may be difficult to find a stable static spherically symmetric solution (i.e., a Schwarzschild analogue) in such theories\cite{dolgov}. In fact Clifton and Barrow showed that in order 
to be consistent with local astronomy, the departure of an $f(R)$ theory from GR should be very small\cite{barrow2}. In a very comprehensive work, Sotiriou and Faraoni\cite{faraoni} reviewed different aspects of $f(R)$ theories.\\

For a local distribution of mass, a gravitational collapse may lead to different possibilities. For example, if the end product of a collapse is a singularity, then the question arises whether the singularity is hidden from the exterior by an event horizon or is visible for an observer. For a systematic study of gravitational collapse we refer to the monograph by Joshi\cite{pankaj}. A brief but recent account of the implications of a gravitational collapse also has been given by Joshi\cite{pankaj1}.  \\

Although various aspects of $f(R)$ gravity has been investigated, the outcome of a collapse has hardly been discussed. There has been a recent investigation using numerical simulations by Borisov, Jain and Zhang\cite{borisov}. Guo, Wang and Frolov\cite{guo} also used a numerical simulation to investigate an $f(R)$ collapse in Einstein frame, where the nonlinear contribution of the curvature in the action is reduced to a nonminimally coupled scalar field via a conformal transformation. Sharif and Kausar\cite{sharif1} investigated the collapse of a spherical fluid distribution with a constant Ricci curvature. Kausar and Noureen\cite{kausar} worked on the effect of an anisotropy and dissipation in the fluid distribution of a collapsing sphere in $f(R)$ gravity. Sharif and Yousaf also studied the stability of the collapsing models in $f(R)$ gravity theories\cite{sharif2, sharif3}. The condition for the validity of a Birkhoff-like theorem in these theories has been discussed by Nzioki, Goswami and Dunsby \cite{rituparno1}.\\

A rigorous account of the gravitational collapse in $f(R)$ gravity has very recently been given by Goswami {\it et al}\cite{rituparno2}. This is an analytical work and has some significant general conclusions. For instance, they showed that for a consistent collapsing model leading to a black hole, one requires an inhomogeneity in the fluid distribution. They choose a widely used form of $f(R)$, given as $f(R) = R + \alpha R^{2}$ to go find an exact solution for a collapsing model.\\

The motivation of the present work is to investigate the possibility of the formation of a black hole or a naked singularity as a result of a perfect fluid collapse in an $f(R)$ gravity model. We assume a simple metric to start with, so in that sense it is not very general, but we find an exact solution for a  collapsing model which is valid for a fair domain of the theory. The density and pressure are inhomogeneous, so the result is consistent with that obtained by Goswami {\it et al}\cite{rituparno2}. The possibility of the formation of a black hole, i.e. an apparent horizon or a naked singularity as the end product of the collapse is found to be dependent on the initial conditions.

The next section deals with the collapsing model. In section 3 and 4, matching with an exterior solution and the visibility of the central singularty is investigated. The 4th and final section includes a discussion on results obtained.

\section{Collapsing model and formation of singularity}

In $f(R)$ theories, the Einstein-Hilbert action of General Relativity is modified by using a general analytic function $f(R)$ instead of $R$. The action is given by

\begin{equation}\label{action}
A=\int\Bigg(\frac{f(R)}{16\pi G}+L_{m}\Bigg)\sqrt{-g}~d^{4}x,
\end{equation}

where $L_{m}$ is the Lagrangian for the matter distribution.

In what follows, we take up the standard metric formulation where the action is varied with respect to $g_{\mu\nu}$ as opposed to a Palatini variation where both of the metric and the affine connections are taken as the arguments of variation. The variation of the action (\ref{action}) with respect to the metric tensor leads to the following fourth order partial differential equation as the field equation,

\begin{equation}
F(R)R_{\mu\nu}-\frac{1}{2}f(R)g_{\mu\nu}-\nabla_{\mu}\nabla_{\nu}F(R)+g_{\mu\nu}\Box{F(R)}=-8\pi G T^{m}_{\mu\nu},
\end{equation}
where $F(R)=\frac{df}{dR}$.

Writing this equation in the form of Einstein tensor, one obtains
\begin{equation}
G_{\mu\nu}=\frac{\kappa}{F}(T^{m}_{\mu\nu}+T^{D}_{\mu\nu}),
\end{equation}
where
\begin{equation}\label{curvstresstensor}
T^{D}_{\mu\nu}=\frac{1}{\kappa}\Bigg(\frac{f(R)-RF(R)}{2}g_{\mu\nu}+\nabla_{\mu}\nabla_{\nu}F(R)-g_{\mu\nu}\Box{F(R)}\Bigg).
\end{equation}

$T^{D}_{\mu\nu}$ represents the contribution of the curvature in addition to Einstein tensor. This may formally be treated as an effective stress-energy tensor $T^{D}_{\mu\nu}$ with a purely geometrical origin. The stress-energy tensor for a perfect fluid is given by $T^{m}_{\mu\nu}=(\rho+p)v_{\mu}v_{\nu}-pg_{\mu\nu}$. Here $\rho$ and $p$ are the density and pressure of the fluid respectively and $v^{\mu}$ is the velocity four-vector of the fluid particles, which, being a timelike vector, can be normalized as $v^{\mu}v_{\mu} = 1$.  \\

The metric is taken to be Lemaitre-Tolman-Bondi type with separable metric components,
\begin{equation}\label{metric}
ds^2=dt^2-B(t)^2X(r)^2dr^2-{r^2B(t)^2}d\Omega^2,
\end{equation}
where $d\Omega^2$ is indeed the metric on a unit two-sphere. \\

The field equations for this metric where the stress-energy tensor is given as sum of $T^{m}_{\mu\nu}$ and $T^{D}_{\mu\nu}$ are given by

\begin{equation}\label{fe1}
3\frac{\dot{B}^2}{B^2}-\frac{1}{B^2X^2}\Bigg[\frac{1}{r^2}-2\frac{X'}{rX}-\frac{X^2}{r^2}\Bigg]=\frac{1}{F}\Bigg[\rho+\frac{f-RF}{2}+\frac{F''}{B^2X^2}+\frac{\dot{B}\dot{F}}{B}+\frac{F'}{B^2X^2}\Big(\frac{2}{r}-\frac{X'}{X}\Big)\Bigg],
\end{equation}

\begin{equation}\label{fe2}
2\frac{\ddot{B}}{B}+\frac{\dot{B}^2}{B^2}-\frac{1}{B^2X^2r^2}+\frac{1}{B^2r^2}=\frac{1}{F}\Bigg[-p+\frac{f-RF}{2}-\ddot{F}+2\frac{\dot{B}\dot{F}}{B}+\frac{2F'}{rX^2B^2}\Bigg],
\end{equation}

\begin{equation}\label{fe3}
\frac{\dot{B}^2}{B^2}+\frac{X'}{rB^2X^3}=\frac{1}{F}\Bigg[-p+\frac{f-RF}{2}-\ddot{F}+\frac{F''}{B^2X^2}+\frac{F'}{B^2X^2}\Big(\frac{1}{r}-\frac{X'}{X}\Big)\Bigg].
\end{equation}
Overhead dot and prime represent differentiation with respect to time $t$ and $r$ respectively.
\\The $G_{01}$ component of the Ricci Tensor gives,
\begin{equation}\label{FR}
\frac{\dot{F'}}{F'}=\frac{\dot{B}}{B}
\end{equation}
which readily integrates to give,
\begin{equation}\label{FandB1}
F'={k_0(r)}B
\end{equation}
where $k_0(r)$ is an arbitrary function of $r$, which comes as a ``constant'' from the integration with respect to $t$. \\
This equation can be written in the form 
\begin{equation}\label{FandB2}
F=B\int{k_0(r)}dr=B(t)k_1(r),
\end{equation}
where we have written $\int{k_0(r)}dr=k_1$. \\

We see that $F(R)$ is also separable as functions of $r$ and $t$. As the Ricci scalar for the metric (\ref{metric}) is of the form $R=R_0 + \psi (r,t)$ (where $R_0$ is a constant), this requirement of separability of $F(R)=\frac{dF}{dR}$ may be difficult to impose for any general functional form of $f(R)$ without any restrictions or special cases. For instance, for $f(R) = R + \alpha R^{2}$, one has to choose the parameter $\alpha$ and the constant $R_0$ such that $1+ 2\alpha R_0=0$. In that sense, this analytical model works alongwith certain restrictions. 

Equations (\ref{fe2}), (\ref{fe3}) and (\ref{FandB2}) can be combined to yield
\begin{equation}\label{eqforB} 
2\frac{\ddot{B}}{B}-2\frac{\dot{B}^2}{B^2}=\frac{1}{B^2}\Bigg(\frac{1}{X^2r^2}-\frac{1}{r^2}+\frac{X'}{rX^3}+\frac{k_1'}{rk_1X^2}-\frac{k_1''}{k_1X^2}+\frac{k_1'X'}{k_1X^3}\Bigg).
\end{equation}
Multiplying both sides by $B^2$, one can easily see that LHS of the resulting equation is a function of time whereas RHS is a function of r only. Therefore both sides must be equal to a constant. Since we are mainly interested in the time evolution of the collapsing system, we concentrate on the time dependent part of the equation,
\begin{equation}\label{b1}
2\frac{\ddot{B}}{B}-2\frac{\dot{B}^2}{B^2}+\frac{\lambda}{B^2}=0,
\end{equation}
where $\lambda$ is the separation constant and is positive. \\
This yields a first integral as
\begin{equation}\label{b2}
\dot{B}^2={\beta}B^2+\frac{\lambda}{2},
\end{equation}
where $\beta$ is a constant of integration. Since we are interested in a collapsing situation we shall henceforth be using the negative root, i.e., $\dot{B}<0$.
With this assumption equation (\ref{b2}) is integrated to yield a simple solution.
\begin{equation}\label{b-int}
B(t)=\frac{1}{2}e^{\sqrt{\beta}(t_0-t)}-\frac{\lambda}{4\beta}e^{-\sqrt{\beta}(t_0-t)}.
\end{equation}

The $r$-dependent part of equation (\ref{eqforB}) gives a relation between $k_1(r)$ and $X(r)$ as
\begin{equation}\label{k1-X}
\frac{1}{X^2r^2}-\frac{1}{r^2}+\frac{X'}{rX^3}+\frac{k_1'}{rk_1X^2}-\frac{k_1''}{k_1X^2}+\frac{k_1'X'}{k_1X^3}+\lambda=0.
\end{equation}

Using equation (\ref{b2}) in equations (\ref{fe1})and (\ref{fe2}), we write the expressions of density and pressure in terms of $f(R)$ and the metric coefficients as
\begin{equation}\label{density}
\rho=-(4\beta k_1+\frac{f}{2F})B-\frac{k_1}{B}\Bigg(2\lambda+\frac{2}{r^2}+\frac{k_1''}{k_1X^2}+\frac{2}{r^2X^2}+\frac{2k_1'}{rk_1X^2}-\frac{X'k_1'}{k_1X^3}\Bigg),
\end{equation}
\begin{equation}\label{pressure}
p=(4\beta k_1+\frac{f}{2F})B+\frac{k_1}{B}\Bigg(\frac{7\lambda}{2}+\frac{2}{r^2X^2}+\frac{2X'}{rX^3}+\frac{2k_1'}{rk_1X^2}\Bigg).
\end{equation}

Using the definition of Misner-Sharp mass function, which actually gives the total energy contained by the sphere\cite{sharp},
\begin{equation}\label{misner}
m(r,t)=\Bigg[\frac{C}{2}(1+\frac{\dot{C}^2}{A^2}-\frac{C'^2}{B^2})\Bigg], 
\end{equation}
for a general spherically symmetric spacetime given by $ds^2=A^2(t,r)dt^2-B^2(t,r)dr^2-C^2(t,r)d\Omega^2$. Using the first integral (\ref{b2}) we can write 
\begin{equation}\label{misner1}
m(r,t)=\Bigg[\frac{rB}{2}(1+r^2{\beta}B^2+r^2\frac{\lambda}{2}-\frac{1}{X^2})\Bigg],
\end{equation}
for the present case. \\

One can see from equation (\ref{b-int}) that when $t=t_0-\frac{1}{2\sqrt{\beta}}\ln(\frac{\lambda}{2\beta})$, $B(t)$ goes to zero, hence the collapsing fluid crushes to a singularity of zero proper volume whose measure is $\sqrt{-g} = B^{3}X$. Equations (\ref{density}), (\ref{pressure}) show that both the fluid density and pressure diverge to infinitely large values at the singularity. As these physical quantities, like density and pressure, are all functions of $r$, the fluid distribution is not spatially homogeneous. The collapse is thus different from the Oppenheimer-Snyder collapse in general relativity\cite{openheimer}.\\

The Ricci Scalar $R = R^{\alpha}_{\alpha}$ for the metric (\ref{metric}) is given by
\begin{equation}
\label{ricci}
R=-12\beta-\frac{2}{B^2}\Big(3\frac{\lambda}{2}+\frac{2X'}{rX^3}+\frac{1}{r^2}-\frac{1}{r^2X^2}\Big),
\end{equation}
and the Kretschmann scalar is given by
\begin{equation}\label{krets}
K=6\beta^2+\frac{1}{B^4}\Bigg[4\frac{(rX^3\dot{B}^2+X')^2}{r^2X^6}+2\frac{(-1+X^2+r^2X^2\dot{B}^2)^2}{r^2X^4}\Bigg].
\end{equation}

One can easily note that both these scalars blow up to infinity at $t=t_{s}$ where $B\Rightarrow 0$. Thus this is indeed a curvature singularity.  \\ 

At the singularity $g_{\theta\theta} = 0$. This ensures that the singularity is a shell focusing singularity and not a shell crossing singularity\cite{yodzis, lake}.

\section{Matching of the collapsing sphere with an exterior vacuum spacetime}
The parameters $\lambda$, $\beta$ and $t_0$ can be estimated from suitable matching of the interior collapsing fluid with a vacuum exterior geometry. Generally, in general relativistic collapsing models, the interior is matched with a vacuum Schwarzschild exterior, which implies continuity of both the metric and the extrinsic curvature on the boundary (\cite{Darmo, Israel}). However, in f(R) theories of gravity, continuity of the Ricci scalar across the boundary surface and continuity of its normal derivative are also required (\cite{Clifton, Deru, Seno}). It has been shown that a Schwarzschild solution is the stable limit of certain $f(R)$ theories\cite{rituparno1}. In connection with the Schwarzchild limit in $f(R)$ gravity we also refer to the recent work by Ganguly {\it et al}\cite{gang}. As the non-Schwarzschild counterparts are obtained mainly for $\frac{1}{R}$ theories which will hardly fall into the scheme of a separable $\frac{df}{dR}$ models, we match our solutions to an exterior Schwarzschild solution.  \\

Matching of the first and second fundamenal form across the boundary hypersurface $\Sigma$ yields:
\begin{equation}
m(t,r) {=^\Sigma} M,
\end{equation}

\begin{equation}
X(r){=^\Sigma} \frac{1}{\Big(1+\lambda\frac{r^2}{4}\Big)^{1/2}}
\end{equation}
where $m(t,r)$ is the Misner-Sharp mass defined by (\ref{misner1}) and $M$ is the Schwarzschild mass.

The problem of matching Ricci Scalar and its' normal derivative was studied in detail by Deruelle, Sasaki and Sendouda (\cite{Deru}). They generalized the Israel junction conditions (\cite{Israel}) for this class of theories by direct integration of the field equations. It was utilised by Clifton et. al. (\cite{Clifton}) and Goswami et. al. (\cite{rituparno2}) quite recently. Following these investigations, the smooth matching of Ricci scalar and its spatial derivative across the boundary hypersurface is discussed in brief. \\

For a spherical geometry where the time-evolution is governed by (\ref{b-int}), a smooth boundary matching of  the Ricci scalar requires that the scalar can be taken in a general functional form 
\begin{equation}
R=T(t)+\frac{f_{1}(r)}{f_{2}(t)}.
\end{equation}
Here $T(t)$, $f_{1}(r)$ and $f_{2}(t)$ are defined in terms of $t$, $r$ and parameters such as $\lambda$, $\beta$ and $\Lambda_1$.
Therefore, at the boundary $r=r_{\Sigma}$, by an inspection of the continuity of $R'$, one can write
\begin{equation}
\frac{2X'}{rX^3}+\frac{1}{r^2}-\frac{1}{r^2X^2}{=^\Sigma} \Lambda_1,
\end{equation}
where $\Lambda_1$ is a constant which can be estimated in terms of the initial conditions, e.g. the parameter $\lambda$.

\section{Visibility of the central singularity}

The condition for the formation of an apparent horizon is given by
\begin{equation}
g^{\mu\nu}Y,_{\mu}Y,_{\nu}=0
\end{equation}
where $Y$ is the proper radius of the two-sphere. So $Y = rB(t)$ in the present case. Thus the relevant equation reads as
\begin{equation}\label{apphor1}
{r^2}{\dot{B}}^2-\frac{1}{X^2}=0.
\end{equation}

Taking advantage of the fact that $B$ and $X$ are functions of single variables, namely $t$ and $r$ respectively, one can write,
\begin{equation}\label{apphor2}
\dot{B}^2=\frac{1}{r^2X^2}=\delta^2,
\end{equation}
where $\delta$ is constant. \\

Using equations (\ref{b2}) and (\ref{apphor2}), one can find, by some simple algebra, the time ($t_{ap}$) of formation of the apparent horizon as

\begin{equation}\label{t-app}
t_{ap}=t_0-\frac{1}{\sqrt{\beta}}\ln\Bigg(\sqrt{\frac{\delta^2}{\beta}}{\pm}\sqrt{\frac{\delta^2-\frac{\lambda}{2}}{\beta}\Bigg)}.
\end{equation}

This immediately yields the condition for the formation of the apparent horizon as ${\delta}^{2}\geq \frac{\lambda}{2}$. \\

From equation (\ref{b-int}), the time ($t_s$) of formation of singularity ($B=0$) is given by
\begin{equation}\label{t-s}
t_s=t_0-\frac{1}{2\sqrt{\beta}}\ln(\frac{\lambda}{2\beta}).
\end{equation}

Depending on $\lambda$, and $\delta$, the visibilty of the central singularity is determined. From the last two equations, 
one has
\begin{equation}
\label{t-app}
t_{s} - t_{ap} = \frac{1}{\sqrt{\beta}} \ln \Bigg[\frac{(\delta {\pm} \sqrt{{\delta}^{2} - \frac{\lambda}{2}})}{\sqrt{\lambda}}\Bigg].
\end{equation}

As this singularities are independent of $r$, this scenario is essentially a non-central one and appears at all points simultaneously. It was shown by Joshi, Goswami and Dadhich \cite{naresh} that in such a case, there is no possibility of a naked singularity. From equation (\ref{t-app}), it is easily found that the consistency with this result impose the boundary condition ${\delta}^2 > \frac{3\lambda}{4}$ . This also satisfies the condition for a real solution of the equation (\ref{t-app}) in order to ensure the appearance of an apparent horizon.

\section{Discussion}

With a simple metric where the metric components are separable as products of functions of time and the radial coordinate, a spherically symmetric gravitational collapse in a framework of $f(R)$ theories of gravity is discussed in the present work. $\frac{df}{dR}$ also happens to be a separable function of $r$ and $t$ coordinates which defines certain conditional domains for this theory. It is shown that the collapse necessarily leads to a singularity of zero proper volume, and the physical quantities like density, pressure etc. all diverge to infinity. The question of the formation of apparent horizon depends on the relative values of $\lambda$ and $\delta$, both of which are separation constants. These constants may be fixed either by matching the collapsing solution to the exterior metric or by initial conditions. In $f(R)$ gravity, a stable Schwarzschild analogue is not guaranteed\cite{dolgov}. However, there are examples of such an analogue for quite a general class of $f(R)$ theories. Assuming the existence of a stable Schwarzchild solution for the exterior, matching at the boundary is discussed. \\
It deserves mention at this stage that one can find the condition for a vacuum collapse by simply setting $p=\rho=0$. This puts a condition 
\begin{equation}
 k_1(r)=\frac{r}{X}e^{\int\Big(\frac{1}{r}-\frac{\lambda r}{2}\Big)X^2dr},
\end{equation}
which is a result of the simplication of the expressions for the density and pressure in equations (\ref{density}, \ref{pressure}).

The conclusion that the the density and pressure remaining inhomogeneous strongly supports the result obtained by Goswami {\it et al}, which is the only extensive work in $f(R)$ collapse. The advantage of the present model is that this is a simple solution, and thus could be useful for any further study. In particular, this simple models can potentially serve a secondary purpose. There is no significant knowledge regarding the possible clustering of dark energy, it is more or less granted that it does not cluster at any scale smaller than the Hubble scale. The investigation regarding collapse may also indicate the possibilities in this connection in a modified theory of gravity.

\end{document}